\documentclass[12pt,a4paper]{article}
\usepackage[top=1.2in, left=0.8in, right=0.8in, bottom=1.2in]{geometry}
\usepackage[numbers]{natbib}
\usepackage{booktabs} 
\usepackage{textgreek} 
\usepackage[ruled]{algorithm2e} 
\usepackage{algorithmic}		
\usepackage{amsmath} 
\usepackage{multirow}
\usepackage{graphicx}
\usepackage{subcaption}
\usepackage{float}
\usepackage{txfonts}	
\usepackage[T1]{fontenc}
\usepackage[colorlinks=true,linkcolor=blue,citecolor=magenta,urlcolor=red]{hyperref}

\def\enabler{{\Large\textepsilon }n\textalpha bler}
\newcommand{\R}{\mathbb{R}}

\newcommand{\footremember}[2]{%
    \footnote{#2}
    \newcounter{#1}
    \setcounter{#1}{\value{footnote}}%
}
\newcommand{\footrecall}[1]{%
    \footnotemark[\value{#1}]%
} 

\author{%
   Evripides Tzamousis  \footremember{trick}{University of Crete and Foundation for Research and Technology-Hellas, Greece.}
   \and Maria Papadopouli \footrecall{trick} \footremember{second trick}{Contact author: Maria Papadopouli (\texttt{mgp@ics.forth.gr)}.} 
  }
\date{}

\begin{document}
\title{On hybrid modular recommendation systems for video streaming}
\maketitle
\begin{footnotesize}
\noindent
{\bf Abstract}
The technological advances in networking, mobile computing, and systems, have triggered a dramatic increase in content delivery services. This massive availability of multimedia content and the tight time constraints during searching for the appropriate content,  impose various requirements towards maintaining the user engagement. The recommendation systems address this problem by recommending appropriate personalized content to users, exploiting information about their preferences. A plethora of recommendation algorithms has been introduced. However the selection of the best recommendation algorithm in the context of a specific service is challenging. Depending on the input, the performance of the recommendation algorithms varies. To address this issue, hybrid recommendation systems have been proposed, aiming to improve the accuracy by efficiently combining several recommendation algorithms. This paper proposes the {\Large\textepsilon }n\textalpha bler, a hybrid recommendation system which employs various machine-learning (ML) algorithms for learning an efficient combination of a diverse set of recommendation algorithms and selects the best blending for a given input. Specifically, it integrates three main layers, namely, the trainer which trains the underlying recommenders, the blender which determines the most efficient combination of the recommenders, and the tester for assessing the system's performance. The {\Large\textepsilon }n\textalpha bler incorporates a variety of recommendation algorithms that span from collaborative filtering and content-based techniques to ones based on neural networks. The {\Large\textepsilon }n\textalpha bler uses the nested cross-validation for automatically selecting the best ML algorithm along with its hyper-parameter values for the given input, according to a specific metric, avoiding optimistic estimation.  Due to its modularity, it can be easily extended to include other recommenders and blenders. The {\Large\textepsilon }n\textalpha bler has been extensively evaluated in the context of video-streaming services. It outperforms various other algorithms, when tested on the "Movielens 1M" benchmark dataset. For example, it achieves an RMSE of 0.8206, compared to the state-of-the-art performance of the AutoRec \cite{sedhain2015autorec} and SVD \cite{koren2009matrix}, 0.827 and 0.845, respectively. A pilot web-based recommendation system was developed and tested in the production environment of a large telecom operator in Greece. Volunteer customers of the video-streaming service provided by the telecom operator employed the system in the context of an "out-in-the-wild" field study. The outcome of this field study was encouraging. For example, the viewing session of recommended content is longer compared to the one selected from other sources. Moreover an offline post-analysis of the {\Large\textepsilon }n\textalpha bler, using the collected ratings of the pilot, demonstrated that it significantly outperforms several popular recommendation algorithms, such as the SVD, exhibiting an RMSE improvement higher than 16\%.

\end{footnotesize}

\def\enabler{{\LARGE\textepsilon}n\textalpha bler}
\section{Introduction}
The dramatic increase of multimedia content along with the plethora of options for content services impose several challenges in maintaining the user engagement in the context of a specific service. The recommendation systems (RS) address this problem through accurately and efficiently recommending users the appropriate content, by exploiting information about their preferences.
In the past two decades, several recommendation algorithms have been developed. They can be classified into two main categories, namely, the content-based and the collaborative filtering (CF) algorithms. The content-based techniques employ the attributes that describe the items (e.g., genres of movies), in order to build profiles of user interests \cite{lops2011content}. On the other hand, the collaborative filtering algorithms utilize historical information about the user preferences (e.g., ratings) to identify similar users or similar items, based on the user rating patterns. The most widely-employed collaborative filtering algorithms are variations of the k-nearest neighbors (k-NN) approach (e.g., user-based and item-based CF) \cite{sarwar2001item,konstan1997grouplens,deshpande2004item,adamopoulos2014over}, as well as various matrix factorization techniques (e.g., SVD) \cite{koren2009matrix,koren2008factorization,lawrence2009non,weston2013nonlinear}. Recently, (deep) neural network methods for recommendation tasks have started to emerge, forming a new subclass of recommendation algorithms \cite{sedhain2015autorec,strub2016hybrid,zheng2016neural}.
\par
Selecting the best recommendation algorithm for a given service is not an easy task: depending on the input, the performance of the recommendation algorithms varies \cite{herlocker2004evaluating,adomavicius2012impact,ekstrand2012recommenders}. For example, although in general collaborative filtering techniques perform better than content-based ones, their prediction accuracy is relatively low, when there is insufficient number of ratings \cite{kluver2014evaluating}, while content-based techniques can alleviate this sparsity problem \cite{henriques2016combining}. To address these limitations, {hybrid} RS, which integrate different recommendation algorithms, have been developed aiming to harness their strengths to provide more accurate predictions than each individual algorithm separately \cite{burke2002hybrid}. \textit{Stacking} or \textit{blending} \cite{wolpert1992stacked} is a common hybridization scheme, which combines predictions derived from the individual recommendation algorithms in order to obtain a final prediction. The appropriate combination of these predictions is based on supervised machine learning (ML) algorithms (e.g., linear regression, artificial neural networks) that assign to each individual algorithm a weight for controlling their impact on the final prediction \cite{jahrer2010combining}. Moreover, the integration of \textit{meta-features} based on rating information, such as the number of user ratings or the average rating of an item, can improve the efficiency of the recommendation algorithms \cite{sill2009feature,bao2009stacking}.
\par
We developed the \enabler, a hybrid recommendation system which employs various ML algorithms for learning an efficient combination of a diverse set of recommendation algorithms (also referred as \textit{recommenders}) and selects the best blending for a given input. It consists of three main layers, namely, the \textit{trainer} which trains the underlying recommenders, the \textit{blender} which determines the most efficient combination of the recommenders, and the \textit{tester} for assessing the system's performance. The \enabler~incorporates recommendation algorithms that span from collaborative filtering and content-based techniques to recently-introduced neural-based ones. Multiple ML algorithms, namely, the Linear Regression, Artificial Neural Network, and Random Forest perform the blending. Specifically, the \enabler~automatically selects the blending algorithm along with its hyper-parameter values (selected from a set of predefined values) that optimizes a specific metric and reports its performance, given the input. The blending process can be further improved considering meta-features, which correspond to statistics on the ratings of specific users and movies. Through meta-features, the blender incorporates user/item profile information. Due to its modularity, the \enabler~can be easily extended to include other recommendation and blending algorithms. 
\par
The \enabler~has been extensively evaluated in the context of video-streaming services. The \enabler~outperforms other systems when tested using the benchmark dataset "Movielens 1M", achieving a RMSE of 0.8206, compared to the state-of-the-art performance of the AutoRec \cite{sedhain2015autorec} and SVD \cite{koren2009matrix} (0.827, and 0.845, respectively). Furthermore, we developed a web-based pilot RS based on the \enabler 's main functionality for a popular video-streaming service provided by a large telecom operator in Greece. We then performed a field study with volunteer subscribers in the production environment of that provider obtaining encouraging results. For example, the viewing session of recommended content was longer compared to the one selected from other sources. A follow-up offline analysis showed that the \enabler~significantly outperforms individual recommendation algorithm in terms of RMSE (an improvement of 16\% or more). 
\par
To the best of our knowledge, the \enabler~is the first hybrid recommendation system that trains various blenders and automatically selects the best one for a given input, unlike other hybrid recommenders that employ a specific blender. The nested cross-validation for the selection and the evaluation of the best blender aims to avoid the optimistic performance estimation \cite{tsamardinos2014performance}. 
\par
The paper is structured as follows: Section 2 overviews the state-of-the-art on hybrid RS, while Section 3 presents in detail the design of the \enabler. Section 4 focuses on its performance analysis using the ''Movielens 1M'' dataset and Section 5 describes the pilot RS that was developed and analyzes the data collected in the field study. Finally, Section 6 summarizes our conclusions and future work plan. 

\section{Related Work}
The hybrid RS emerged by the need to effectively combine various recommendation techniques in order to improve the recommendation performance \cite{burke2002hybrid}. Given that each recommender captures different aspects of the data and trains its corresponding model in a different way, it manifests different strengths and weaknesses. By their appropriate combination, a hybrid RS can focus on their strengths and thus improving the accuracy \cite{wozniak2014survey}. A range of hybridization techniques has been proposed that require training machine learning models. 
\par 
\textit{Switching}, a simple hybridization technique, selects a single recommendation algorithm among various ones, based on a criterion. Lekakos and Caravelas  \cite{lekakos2008hybrid} propose a hybrid RS that uses the user-based collaborative filtering as the primary recommendation algorithm and shifts to a content-based one, when a given active user (namely the user that the prediction refers to) has not an adequate number of similar users. Instead of employing one recommender at a time, the \textit{mixed} hybridization approach allows the performance of multiple recommenders. This approach is widely-used by Netflix \cite{gomez2016netflix}. The optimal presentation of multiple recommendation lists in a single page is known as the \textit{full page optimization} problem \cite{amatriain2016past}. The \textit{cascade}, a technique largely used by Youtube \cite{covington2016deep}, at first produces a small subset of movies of interest to the user, and then, by employing a separate algorithm, aims at distinguishing a few ''best'' recommendations. The hybridization approaches are not limited to the above: a more detailed description can be found in \cite{burke2002hybrid,wozniak2014survey}. 
\par 
{Blending} is a common hybridization approach,  also used by the \enabler. It gained substantial attention during the Netflix Prize competition, where it had a key contribution to the winning solution. In a hybrid RS that employs blending, multiple recommendation algorithms are trained, each one of them producing predictions separately. At a later stage, the \textit{blending} algorithm is trained to appropriately combine the predictions of the different algorithms, using their prediction scores as features, in order to improve the final prediction accuracy. The teams were required to build a RS that would minimize the root-mean-square error (RMSE) on a given dataset by 10\%. One of the winning teams employed a linear combination of multiple collaborative filtering algorithms, mostly including variations of k-nearest neighbors and matrix factorization, where the weights of each algorithm are learned by linear regression \cite{bell2008bellkor} . A simpler hybrid RS \cite{chikhaoui2011improved} linearly combines a collaborative filtering algorithm and a content-based one, along with user demographic information. The contribution of each algorithm in each user's final prediction is based on a function that takes into account the number of user ratings, instead of employing a learning procedure that would determine those weights. Dooms \textit{et al.} \cite{dooms2015offline} introduces a user-specific hybrid system that results in different combinations of up to 10 recommenders for each user. The weights of each individual algorithm are determined through an iterative binary search process that adjusts each algorithm's contribution at each step, towards minimization of the final prediction error. It also employs a switching technique that selects the best performing algorithm for each user, although their experiments show that the linear combination of multiple algorithms yields better results.

\par
Subsequent experiments on the Netflix dataset showed that further improvements in the accuracy can be achieved by employing sophisticated blending algorithms that can identify \textit{non-linear} relationships between the individual algorithms. Jahrer \textit{et al.} \cite{jahrer2010combining} tested various blending algorithms, such as linear, kernel, polynomial regression, bagged decision trees and artificial neural networks, and showed that an ANN network with a single hidden layer achieves the lowest RMSE on the Netflix dataset. An advantage of using ANNs for blending is their excellent accuracy and fast prediction time (a few seconds for thousands of testing samples), at the cost of a long training time (hours for millions of training examples).
\par

Although the combination of different recommenders helps alleviating the drawbacks associated with the use of each one  individually, it does not take into account the factors that may affect their individual performance. Adomavicius \textit{et al.} \cite{adomavicius2012impact} investigated the impact of the data characteristics on the performance of several popular recommenders and indicated that the prediction accuracy depends on various factors, such as the number of users, items, and ratings. This means that the suitability of a recommendation algorithm may vary depending on the underlying conditions of the data samples. For example, despite the fact that in general collaborative filtering algorithms perform better than content-based ones, the collaborative filtering ones perform poorly for users or items without a sufficient number of ratings. On the other hand, although content-based approaches address the problem of sparsity in the number of ratings, their performance relies heavily on the available information about the item contents. As a result, to combine more effectively the various recommendation algorithms, the blending algorithms should take into account the additional information that represent properties of the input, known as \textit{meta-features}. The number of ratings from a specific user or about a specific item are prominent meta-features. Stream \cite{bao2009stacking}, a hybrid RS, makes use of meta-features by combining a user-based, an item-based algorithm, and a content-based algorithm that estimates similarities between movies based on their content description. Its bagged model trees algorithm for blending exhibits better prediction accuracy than using linear regression, since the first is capable of identifying more complex non-linear relationships among individual recommenders as well as meta-features. Finally, the feature-weighted linear stacking (FWLS) \cite{sill2009feature} allows for more elaborate combinations of recommenders by linearly combining the predictions of individual recommenders, with their associated coefficients to be parametrized as linear functions of the meta-features.

\section{The {\Large\textepsilon }n\textalpha bler system architecture}
The \enabler~aims to accurately determine the user preferences and provide efficient personalized recommendations in the context of video streaming. It follows a client-server architecture. The client runs on the user's device, where the video-streaming application operates. It collects information about the user preferences and watching behavior and uploads it on the server. The server produces personalized recommendations by applying a set of recommendation algorithms on the collected user information. The predicted ratings of each algorithm are ranked in descending order, so that the content predicted as the most preferred, appears at the top. The generated recommendation lists are sent to the clients for presenting the appropriate content to users. 
\par
\begin{figure}[t!]
\centering
\includegraphics[width=6.7in]{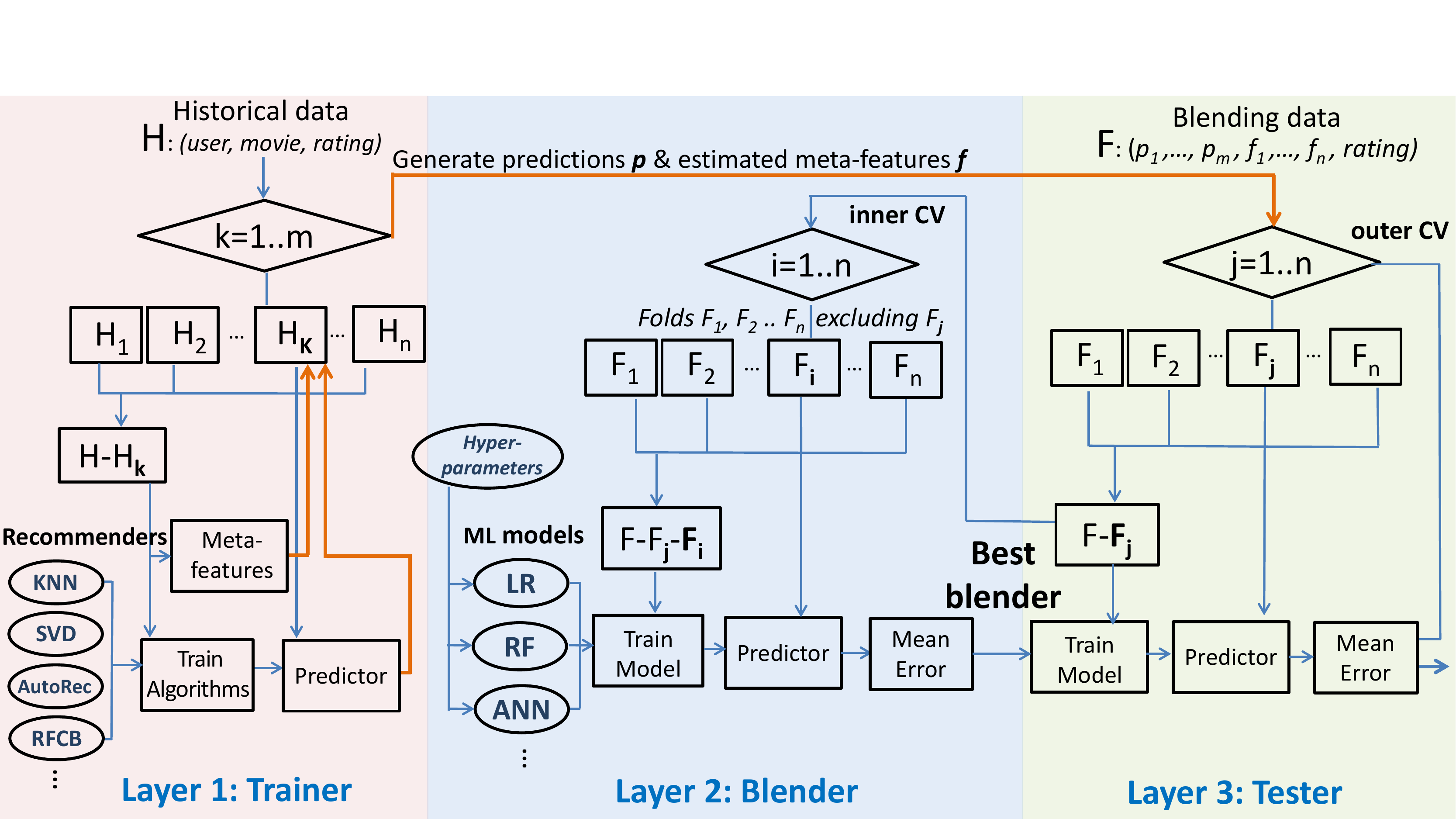}
\caption{The architecture of the \enabler~server with its three main layers, namely, the \textit{trainer}, \textit{blender}, and \textit{tester}. The blender takes into consideration various meta-features, e.g., number of ratings from a specific user or about a specific item. The blending set includes the actual ratings provided by users for movies and the predictions of the different recommenders (e.g., $p_1$,..$p_m$) and is formed at the Trainer layer. After the completion of the Trainer, the nested cross-validation of the Blender and Tester follows.}
\label{fig:arch}
\end{figure}

The main layers of the \enabler~server, namely, the \textit{trainer} which trains the underlying recommenders, the \textit{blender} which determines the most efficient combination of the recommenders, and the \textit{tester} for assessing the system's performance are shown in Fig.~\ref{fig:arch}.     
The \enabler~server aims to identify which blending algorithm results in the most appropriate combination of the underlying recommendation algorithms, for optimizing a specific performance metric. The root-mean-square error (RMSE) has been used. However, other performance metrics can also be employed.
The Trainer layer performs a cross-validation. During this process, the blending set is formed. After the completion of the Trainer, the Blender and Tester follow. A part of the blending set will be employed for training the blending model (Blender) and the remaining for assessing its performance (Tester). The following sections describe the process in detail.

\subsection{The \textit{Trainer} layer}
The trainer (layer 1) is responsible for training the individual recommenders using the historical data (H), which consist of user-movie-ratings triplets. Cross-validation (CV) is performed on the historical data, partitioning the data to disjoint folds $H_k$ and forming the set of folds H. In each k-th fold iteration, each recommender is trained on the $H-H_k$ folds, which serve as the training set, and estimate the predictions for the user-movie pairs found in $H_k$. Statistics on the meta-features are also estimated for the user-movie pairs in $H_k$, taking into account the subset of the historical data that corresponds to each iteration's training set $(H-H_k)$ folds. Let as assume that \enabler~employs $m$ recommenders and $n$ meta-features for users and movies. For each user-movie input, the \enabler~reports the following predictions ($p_1,..., p_m$) derived by the underlying recommenders and the values of the meta-features ($f_1,..., f_n$) that have been calculated during the aforementioned CV procedure, forming the blending dataset (also referred as \textit{blendset}). The blendset will then be used for training the various blending models, with $(p_1,..., p_m , f_1,..., f_n)$ being the features that are given as input to the ML algorithms as shown in Figure ~\ref{fig:arch}.
\par
The \enabler~incorporates a diverse set of recommenders, namely two popular k-NN algorithms, the SVD, the AutoRec which achieves state-of-the-art accuracy, and a CB technique. 

\par\par
The \textbf{user-based collaborative filtering} (UBCF) \cite{resnick1994grouplens} is based on the hypothesis that similar users have similar preferences for similar items. A similarity function $sim(u,v)$ computes the level of agreement between users. A {neighborhood} $G$ is built for a user $u$, consisting of the set of users who are similar to $u$ (also referred as \textit{neighbors} of $u$). Once $G$ has been determined, the prediction for the user $u$ about the item $i$ $p_{u,i}$ is the weighted average over the ratings of his/her neighbors on this item, using as weights the computed similarities:
\begin{equation}
p_{u,i} = \bar{r_{u}} + \frac{\sum_{v\in G(u)}sim(u,v)(r_{v,i}-\bar{r_{v}})}
{ \sum_{v\in G(u)}|sim(u,v)| } 
\end{equation}
where $G(u)$ is the neighborhood of the user $u$ and $v$ is a neighbor of $u$ who has rated the item $i$. The variables $\bar{r_{u}}$ and $\bar{r_{v}}$ correspond to the average rating of the user $u$ and $v$, respectively. The size of the neighborhood affects the final accuracy of the algorithm. A neighborhood size in the range of 20 to 50 is recommended \cite{herlocker2004evaluating,ekstrand2011rethinking}. 

\par\par
The \textbf{item-based collaborative filtering} (IBCF) algorithm relies on the idea that the difference in the ratings of similar items provided by a given user is relatively low \cite{sarwar2001item,linden2003amazon}, and aims at computing similarities between items, instead of users. Two items are considered similar if the same users tend to rate them similarly. Finally, the expected rating of the user $u$ for the movie $i$ $p_(u,i)$ is estimated as follows:
\begin{equation}
p_{u,i} = \frac{\sum_{j\in G(i)}sim(i,j)(r_{u,j})}
{ \sum_{j\in G(i)}|sim(i,j)| } 
\end{equation}
where $G(i)$ is the neighborhood of the item $i$ and $j$ is a neighbor of $i$ that the user $u$ has rated with a value of $r_{u,j}$. Experiments on various datasets indicate that in general the IBCF performs better than the UBCF in terms of accuracy \cite{sarwar2001item, ekstrand2011rethinking}. To avoid introducing noise, excluding the items with low similarity from the neighborhood is recommended \cite{sarwar2001item}.

\par \par
The \textbf{SVD} \cite{koren2009matrix} uses an unsupervised learning method for decomposing the matrix $R$ that contains the ratings provided by users to movies into a product of two other matrices $P$ and $Q$ of a lower dimensionality $F$. Each user $u$ (movie $i$) is represented by a vector $p_u$ ($q_i$) that corresponds to $F$ latent features, describing each user (movie), respectively. Suppose that $M$ is the number of users and $N$ the number of movies. These vectors form the two matrices, $P \in \R ^{M \times F}$ and $Q \in \R ^{N \times F}$. The prediction is given by the dot product of the user vector $p_u$ and the item vector $q_i$, as shown in the following equation:
\begin{equation}
p_{u,i} = q_{i}^T p_{u} = \sum_{f =1}^{F} q_{i,f} \cdot p_{u,f}   
\end{equation}
Typically, the initial ratings matrix $R$ is sparse. As a result, the SVD aims at approximating the matrix $R$ by utilizing the discovered latent features of users and items, and producing predictions for the missing values. In the Netflix competition, SVD was shown to be one of the most important class of algorithms and several studies indicate its superior accuracy compared to the UBCF and IBCF.

\par \par
The \textbf {AutoRec} \cite{sedhain2015autorec} is a recently-introduced collaborative filtering algorithm, based on the autoencoder paradigm; an artificial neural network that learns a representation (encoding) of the input data with the aim of reconstructing them in the output layer. The AutoRec takes as input each partially observed movie vector $r_i \in \R ^M$ that consists of the ratings given by users, and those that are unknown, projects it into a low-dimensional space of size K, and then reconstructs it in the output layer, filling the missing values with predictions. The reconstruction $h(r_i;\theta)$ of the input item vector $r_i$ is given by
\begin{equation}
h(r_i;\theta) = f(W\cdot g(Vr + \mu) + b)  
\end{equation}
with $g(\cdot)$ and $f(\cdot)$ being the activation functions employed in the hidden layer and output, respectively.  Here, $\theta$ are the parameters of the AutoRec model, learned using the backpropagation algorithm. More specifically, $V \in \R ^{M\times K}$ and $W \in \R ^{K\times M}$ are the transformation weights for the hidden and output layers, respectively, with their corresponding biases $\mu \in \R ^{K}$ and $b \in \R ^{M}$. Finally, the expected rating of the user $u$ for the movie $i$ is given by
\begin{equation}
p_{u,i} = h(r_i;\theta)_{u}
\end{equation}
Unlike SVD, which embeds both users and items in a latent space, the AutoRec requires learning only item representations. Furthermore, using non-linear activation functions, such as the \textit{sigmoid}, the AutoRec can identify more complex non-linear representations, in contrast to the SVD that focuses on linearities.

\par \par
The \textbf{random forests for content-based} recommendations  (RFCB) employ a random forest approach \cite{breiman2001random} to construct decision trees for each user, using the attributes of the movies they have rated, which form the user profile. The user profile encapsulates the level of preference for each item's attribute. We employ the movie genres for describing each item, but additional attributes can be incorporated. The predicted rating for the user \textit{u} about the item \textit{i} is obtained by averaging out the predictions of each decision tree in the user's profile:
\begin{equation}
p_{u,i} = \sum_{t=1}^{T} \frac{p_{u}^{t}(i)}{T}
\end{equation}
where $p_{u}^{t}(\cdot)$ indicates the prediction of the $t$-th decision tree about the item $i$ and $T$ is the total number of decision trees of user $u$.

\begin{algorithm}[t!]
\small
\caption{Nested Cross-Validation}
\label{alg:nestedcv}
\begin{algorithmic}
\STATE \textbf{Input:} Dataset F, integer k, all possible models $a \in A$ based on the given ML algorithms and their corresponding hyper-parameters 
\STATE \textbf{Output:} Root-mean-square error
\STATE
\STATE partition $F$ randomly into $k$ equal-sized disjoint subsets $F_i$
\FOR{$j$ = 1 to $k$}
\STATE $F^{-j} = F - F_j$
\FOR{each a $\in$ A}
\FOR{all $i$}
\STATE $F^{-j,i} = F^{-j} - F_i$
\STATE train model a on  $F^{-j,i}$  and
\STATE evaluate its performance on $F_i$, obtaining the performance $P_i^{a}$ 
\ENDFOR
\STATE $\bar{P}_{j}^{a}$ = mean($P_i^{a}$, $\forall i$)
\ENDFOR
\STATE select $\hat{a}_{j}$ = $argmax_{a}\bar{P}_{j}^{a}$
\STATE train the model $\hat{a}_{j}$ on $F^{-j}$, and
\STATE evaluate its performance on $F_j$ , obtaining the performance $P_j$
\ENDFOR
\RETURN mean($P_j$, $\forall j$)
\STATE
\end{algorithmic}
\end{algorithm}

\subsection{The \textit{Blender} and \textit{Tester} layers}
The \enabler~currently employs the Linear Regression, Neural Networks, and Random Forest algorithms for building the blending models but can be easily extended to integrate others. Each algorithm has a set of hyper-parameters that can significantly affect the complexity and performance of the models. For determining the most appropriate values for their hyper-parameters and evaluating the performance of each blender, we employ the Nested Cross-Validation (Nested CV), presented in detail in Algorithm 1. The nested CV selects the best performing blending algorithm along with its corresponding hyper-parameter values from a set of predefined values. It then estimates the performance of the final model. More specifically, the tester layer obtains as input the blendset, generated from the trainer layer, and partitions it into equally-sized disjoint folds $F_j$. The blendset is formed at the completion of the cross-validation of the Trainer layer. The nested cross-validation of the Blender and Tester follows the Trainer.
\par
The blender takes as input the training set of the tester's loop $(F - F_{j})$, cross-validates it (i.e., "inner CV"), and reports the best blending model. In the $i$-th iteration of the inner CV loop, the various blending algorithms are trained on the $F-F_j-F_i$ folds with different values on their corresponding hyper-parameters, and, then tested on the $F_i$ fold. The \enabler~evaluates the performance of the best model in each fold iteration (referred as the "outer CV") and reports the mean error for a given dataset. The following paragraphs present the ML algorithms that it currently employs.\par


The \textbf{Artificial Neural Networks (ANNs)} \cite{mitchell1997artificial} consist of an interconnected group of artificial neurons, where the connections are associated with weights, and model complex relationships between the input and the output. A typical structure for an ANN comprises of three types of layers, namely, the input layer that corresponds to the input features (i.e., predictions of individual algorithms and meta-features), the hidden layer that tries to extract patterns and detect complex relations among the features, and the output layer which emits the final prediction score. The weights are adjusted (learned) during the training phase by the backpropagation learning algorithm. The ANNs can efficiently cope with that task of blending different recommendation algorithms \cite{jahrer2010combining}. The number of hidden layers and number of neurons in each hidden layer are hyper-parameters that need to be tuned, resulting in different network architectures.\par

\textbf{The Random Forest (RF)} constructs a multitude of decision trees and outputs as a final prediction the mean prediction of the individual trees \cite{breiman2001random}. Each tree is trained on a random subset of the training data. Let us support that L input features are available for each training example and that at each node $l$ features (with $l << L$) are randomly selected. In our implementation, we employ the Gini impurity metric for measuring the \textit{best} split and use the value of $\sqrt[]{L}$ for the hyper-parameter $l$. In our case, the features correspond to the predictions of the recommenders along with the estimated meta-features. Consequently, each individual tree results in a different combination of the features and encapsulates different relationships among the individual recommenders and meta-features. A random forest model can identify non-linear associations between the input features, which can improve the efficiency the final combination. By increasing the number of the trees, the prediction error decreases, however, albeit with a computation cost. As a result, the number of decision trees that are trained is the hyper-parameter that requires tuning.\par

The \textbf{ Linear Regression (LR)} aims at identifying appropriate weights for the individual recommenders in order to linearly combine them and obtain a final prediction score. Compared to neural networks and random forests, the linear regression can faster determine the coefficients for each recommender. However, it is unable to discover complex nonlinear relationships among the recommenders. Additionally, the weight assigned to each recommender remains unchanged to the different user (movie) characteristics, such as the number of ratings of a user (movie), referred as user (movie) support, respectively. Note that in the case of neural networks and random forests, weights can be different depending on the user/movie characteristics. For addressing this limitation, Jahrer \textit{et al.} \cite{jahrer2010combining} proposed a "binned" version of linear regression, where the dataset is divided into equally sized bins based on a single criterion (e.g., user support). For each bin, a different linear regression model is learned. The linear regression can now account for differences in the performance of the various recommendation algorithms and weight them differently depending on the conditions. The regularization parameter $\lambda$ is a hyper-parameter that needs to be tuned. We experimented with different numbers of bins and try out two discrete criteria for binning.

\subsection{Computational complexity}
The \enabler~consists of two phases, namely the \textit{training} phase and the \textit{runtime} one. The training phase is responsible for training the recommenders and blenders. It then chooses the best performing blender and estimates its performance. This phase, performed offline, is of relatively high computational complexity. It also runs periodically to take into account new user information. To determine when a re-training is required, we record the accuracy over time, along with the performance of each recommender. When the performance falls below a specified threshold, the recommenders are re-trained, and a new blender is determined. Parallelizing the training of the underlying recommendation algorithms can speedup the process. 
\par
The runtime phase of the \enabler~is of negligible computational complexity compared to the training phase. The recommenders can calculate hundreds of thousands of predictions within a few minutes \cite{guo2015librec}. The blending task for computing the final predictions is also computationally inexpensive. More specifically, the runtime phase of linear regression and artificial neural network algorithms is of $O(p)$ ($p$ is the number of recommenders combined), and random forest $O(t\cdot log(e) )$ ($e$ is the total examples in the training phase and $t$ the number of trees employed) \cite{louppe2014understanding}.

\section{Performance Analysis using MovieLens}
The \enabler~was evaluated in terms of root-mean-square error (RMSE) and compared with widely-employed and state-of-the-art recommendation algorithms. The RMSE is the square root of the mean of the squared differences between the predicted and real ratings. Its performance has been also compared with the individual recommendation algorithms that it employs. Moreover, each blending algorithm was separately assessed with various hyper-parameter values to examine how the performance varies. The analysis uses the real-world benchmark dataset Movielens 1M \cite{harper2016movielens}. This dataset contains 1,000,209 ratings of approximately 3,900 movies made by 6,040 users. Each user has provided at least 20 ratings, and each movie has been rated at least once. The ratings are in a scale from 1 to 5.

\begin{figure}[ht]
\centering
\includegraphics[width=3.8in,height=4.3cm]{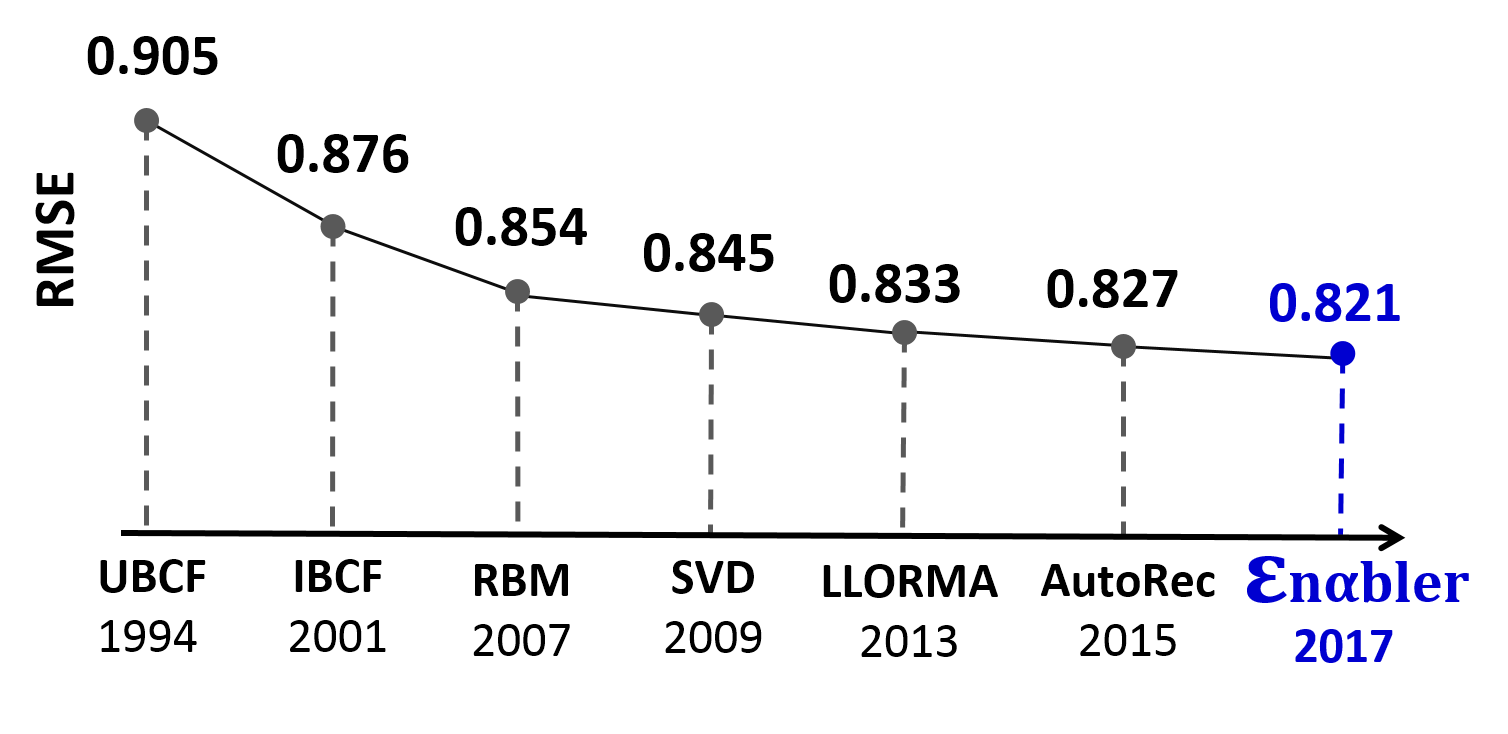}
\caption{The performance of widely-employed and state-of-the-art recommenders, as reported in \cite{sedhain2015autorec} and \cite{guo2015librec}, compared to the \enabler using the Movielens 1M benchmark dataset.}
\label{fig:state_enabler}
\end{figure}

\begin{table}[!ht]
\small
\renewcommand{\arraystretch}{1.3}
\caption{Accuracy of each recommender for specific hyper-parameter values using the Movielens 1M dataset}
\label{table:t1}
\centering
\begin{tabular}{|c|c|c|}
\hline
\multicolumn{2}{|c|}{\textbf{Recommender}} & \textbf{RMSE}\\
\hline
\multirow{2}{*}{User-based collaborative filtering} & {20 neighbors} & {0.9707}\\\cline{2-3} 
                      								& {80 neighbors} & {0.9067}\\\cline{2-3}
\hline
\multirow{2}{*}{Item-based collaborative filtering} & {20 neighbors} & {0.8737}\\\cline{2-3} 
                      								& {80 neighbors} & {0.8763}\\\cline{2-3}
\hline
\multirow{2}{*}{SVD} & {50 factors} & {0.8737}\\\cline{2-3} 
                     & {500 factors} & {0.8496}\\\cline{2-3}
\hline
\multirow{2}{*}{AutoRec} & {100 nodes} & {0.8578}\\\cline{2-3} 
                     	& {300 nodes} & {0.8402}\\\cline{2-3}
\hline
\multirow{2}{*}{Random forests content-based} & {20 trees} & {0.9095}\\\cline{2-3} 
                     						& {30 trees} & {0.8951}\\\cline{2-3}
\hline

\end{tabular}
\end{table}

\subsection{Evaluation setup}
The diverse set of recommenders that was integrated in the \enabler~is shown in Table~\ref{table:t1}. Note that two differently parameterized instances of each recommender were employed to further improve the accuracy of the system \cite{wozniak2014survey}. Toscher \textit{et al.} \cite{toscher2009bigchaos} and Koren \textit{et al.} \cite{koren2009bellkor} also blend multiple instances of the same algorithms and show their added benefit in the final prediction accuracy. The accuracy of each recommender was assessed with the employment of 5-fold CV. Each recommender was trained using 4 folds of samples and tested on the remaining fold, repeating this procedure 5 times, so that all folds were used for testing. For evaluating the performance of the \enabler, the 5-fold nested CV is employed, using 5-fold CV for the performance estimation layer, and 4-fold CV for the model selection layer. To investigate the system's performance with each blending algorithm separately, 5-fold nested CV was applied, restricting the model selection layer to employ a single blender in each experiment.
\par
Each blender has a number of hyper parameters that needs to be tuned. At the model selection layer, all the combinations of the different parameters are tested. For the linear regression algorithm, the set of values of the regularization parameter \textlambda~was $\{0.0001, 0.001, 0.01, 0.1, 1, 10\}$. The user and the movie support were employed as the binning criteria, and the number of bins varied within the values $\{1, 2, 4, 8, 12\}$. The size of a random forest was selected among the values $\{10, 50, 100, 250, 500\}$. The number of hidden layers an ANN used, ranged from 1 to 3, and their node size was selected among the values $\{8, 12, 24$\}. However, in the case of 3 hidden layers, the second and third hidden layer were limited to use 8 or 12 nodes. 

\begin{figure} [t]
\renewcommand{\arraystretch}{1.3}
\centering
\includegraphics[width=3.5in]{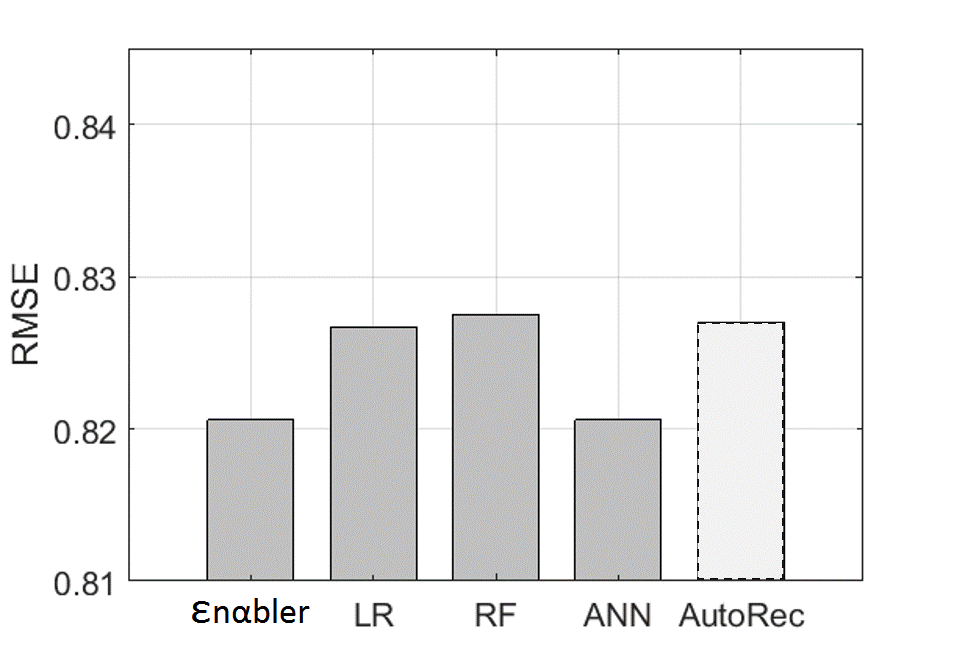}
\footnotesize
\put(-158,123){\scriptsize \setlength{\tabcolsep}{4pt}\begin{tabular}{|c|c|} 
\hline
\textbf{Algorithm} & \textbf{RMSE}\\
\hline
\enabler & 0.8206  \\
\hline
LR  & 0.8267 \\
\hline
RF & 0.8275 \\
\hline
ANN & 0.8206 \\
\hline
AutoRec & 0.827 \\
\hline
\end{tabular}}\\
\caption{RMSE performance of each blender compared to the AutoRec, as reported in \cite{sedhain2015autorec}, using the Movielens 1M benchmark dataset.}
\label{fig:enabler_nested}
\end{figure}

\subsection{Discussion}
\par 
The \enabler~outperforms the AutoRec, achieving an RMSE of 0.821 compared to the 0.827 achieved by the AutoRec, as reported in \cite{sedhain2015autorec} (Fig.~\ref{fig:state_enabler}). The performance of some of the most widely-employed recommendation algorithms, such as the RBM which uses a class of two-layer undirected graphical models \cite{salakhutdinov2007restricted}, and the LLORMA, an SVD-based approach \cite{lee2013local}, is also presented (Fig.~\ref{fig:state_enabler}). Fig.~\ref{fig:state_enabler} depicts the performance of some well-known recommendation algorithms over the years, indicating that the reduction of RMSE by 0.06 achieved by the \enabler, is notable. Let us mention that unlike the deep version of AutoRec with three hidden layers and a number of 500, 250, and 500 nodes, respectively, employed in \cite{sedhain2015autorec}, the \enabler~incorporates a shallow network architecture with a single hidden layer and fewer hidden nodes (Table~\ref{table:t1}). In addition, the AutoRec is evaluated based on a \textit{hold-out} estimation, unlike our approach that employs the nested CV, aiming to derive a more accurate estimate of model prediction performance.
\par
In the following paragraphs, we delve into the performance of each recommender and each blending algorithm, when used in a stand-alone manner.
\par 
\textbf{On the performance of each recommender} Table~\ref{table:t1} shows the recommenders that the \enabler~employed, along with their RMSE performance when used individually. The AutoRec algorithm with a single hidden layer, consisting of 300 hidden nodes, exhibits the best prediction accuracy (0.8402 RMSE) among the various recommenders. The smaller the number of hidden nodes (e.g., 100), the lower the accuracy. For 100 hidden nodes, the accuracy of the AutoRec is lower than of the SVD with 500 and 50 factors (0.8496 and 0.8534), respectively. The UBCF with a small neighborhood size (i.e., 20) has the worst performance, which can be improved as the number of neighbors increases (e.g., 80). On the other hand, the IBCF's accuracy seems to be less sensitive to changes in the neighborhood size, and decreasing the number of neighbor only slightly improves its accuracy. The RFCB has a comparable performance to the best UBCF model.
\par
\textbf{On the performance of each blending algorithm} We furthermore performed extensive experiments with each blending algorithm and their corresponding hyper-parameter values separately to assess its impact on the system's performance. The artificial neural network algorithm achieves the best combination of the recommenders, with a RMSE of 0.8206, followed by the linear regression and random forest algorithms (0.8267, and 0.8275), respectively (Fig.~~\ref{fig:enabler_nested}). All three blending algorithms exhibit a similar or better RMSE performance compared to AutoRec (as presented in Fig.~~\ref{fig:enabler_nested}).

\par

For examining the performance of the linear regression, two criteria for dividing the data were tested, namely, the user, and movie support, and the number of bins varied from 1 to 12 (Table~\ref{table:group} (\textit{left}) summarizes the results). When a single bin is employed, a single linear regression model is trained for all data. For example, when the user support is used for binning, each rating sample is classified in a bin based on the number of ratings the current user has provided. This results in employing different linear regression models for combining the recommenders, depending on the total ratings of the corresponding user. The larger the number of bins, the smaller the RMSE for both criteria (Table~\ref{table:group} (\textit{left})). This indicates that the \textit{binned} version of linear regression can be beneficial in blending various recommendation algorithms. However, the deployment of more than 8 bins seems to lower the prediction accuracy, especially under the user support case (Table~\ref{table:group} (\textit{left})). By increasing the number of bins, we assign less training examples in each bin, making the fitting process of the linear regression more challenging. The best results are obtained for 8 bins using the binning criterion for movie support. It achieves an RMSE of 0.8267, which is slightly better than the AutoRec (Fig.~\ref{fig:enabler_nested}). As shown in Table~\ref{table:group} (\textit{right}), increasing the size of the random forest, improves the prediction accuracy of the system, however with a decreasing benefit. The best performing random forest model performs slightly worse than the AutoRec (0.8275 vs. 0.8270, respectively). The prediction accuracy of different ANN architectures is demonstrated in Table~\ref{table:anns}. The ANN with 3 hidden layers, consisting of 24 neurons in the first layer and 12 neuron in the second and third ones achieves the best RMSE performance (0.8206). By increasing the number of hidden layers, a better combination of the underlying recommenders can be accomplished, resulting in more accurate predictions: as an ANN becomes deeper, it is more capable of identifying complex relations between the different recommendation algorithms, compared to shallow structures \cite{lecun2015deep}. However, even with a single hidden layer, the ANN outperforms both the linear regression and random forest models (0.8232 RMSE with 8 neurons).
\par

\begin{table}[t!]
\centering
\renewcommand{\arraystretch}{1.3}
 \caption{RMSE performance of various Linear regression \textbf{\textit{(left)}}, and Random forest models using the Movielens 1M benchmark dataset\textbf{\textit{(right)}}} 
\footnotesize
\hspace{1cm}
\begin{tabular}[t]{|c|c|c|}
\hline
\textbf{Number of bins} & \textbf{Movie support} & \textbf{User support}\\
\hline
1 & $0.8287$ & $0.8288$ \\
\hline
2 & $0.8277$ & $0.8278$ \\
\hline
4 & $0.8269$ & $0.8270$ \\
\hline
8 & \textbf{0.8267} & \textbf{0.8268}\\
\hline
12 & $0.8268$ & $0.8272$ \\
\hline
\end{tabular}\hfill%
\centering
\begin{tabular}[t]{|c|c|}
  \hline
  \textbf{Number of trees} & \textbf{RMSE}\\
  \hline
  10 & 0.8550  \\
  \hline
  50 & 0.8323  \\
  \hline
  100 & 0.8299 \\
  \hline
  250 & 0.8279 \\
  \hline
  500 & \textbf{0.8275} \\
  \hline
\end{tabular} 
\hspace{2cm}
\label{table:group}
\end{table}

\begin{table}[t!]
\centering
\footnotesize
  \renewcommand{\arraystretch}{1.4}
  \caption{Performance of various ANN models using the Movielens 1M benchmark dataset.} 
  \begin{tabular}[t]{|c|c|c|}
  \hline
  \textbf{Number of hidden layers} & \textbf{Number of hidden nodes} & \textbf{RMSE}\\
  \hline
  \multirow{3}{*}{1} 	& {8}  			& {0.8232}\\\cline{2-3} 
                      & {12} 			& {0.8225}\\\cline{2-3}
                      & {24} 			& {0.8220}\\\hline

  \multirow{2}{*}{2} 	& {24, 12}  	& {0.8218}\\\cline{2-3}
                      & {24, 24}		& {0.8216}\\\hline

  \multirow{2}{*}{3} 	& {24, 12, 8} 	& {0.8208}\\\cline{2-3}
                      & {24, 12, 12} 	& {\textbf{0.8206}}\\\hline
  \end{tabular}
  
\label{table:anns}
\end{table}

\section{Our web-based pilot recommender system}
To assess the performance of personalized recommendation systems in real-world, we developed a web-based pilot RS for providing personalized recommendations to the customers of a major Greek telecom operator for a mobile video-streaming service. Actually, the design and the development of this system occurred prior to the completion of the \enabler. The pilot system implements only partially the \enabler:  its server uses the SVD and IBCF, without employing the blender and tester layers  (Fig.~\ref{fig:arch}). The client runs on Android device, where the video-streaming application operates. Through the application, users can watch live-streamed (LiveTV) video content, such as movies and TV shows, or Video on demand. 
\par
\begin{figure}[h!]
\captionsetup[subfigure]{labelformat=empty}
\centering
\begin{subfigure}[]{0.62\textwidth}
   \includegraphics[width=1\linewidth, trim= 10mm 70mm 12mm 65mm]{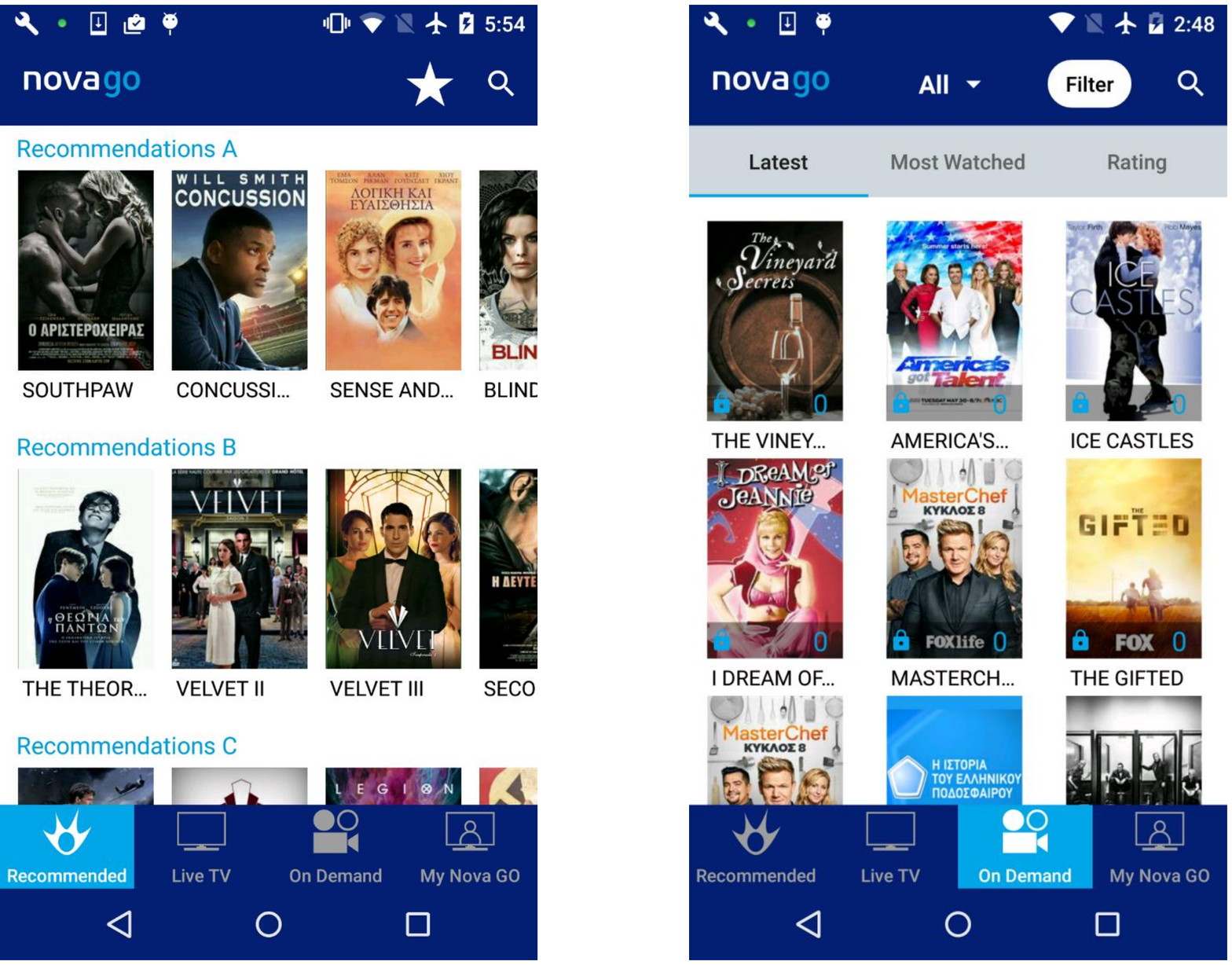}
   \caption{(a) \space\space\space\space\space\space\space\space 
   				\space\space\space\space\space\space\space\space 
                \space\space\space\space\space\space\space\space 
   				\space\space\space\space\space\space\space\space 
                \space\space\space\space\space\space\space\space
                \space\space\space\space\space\space\space\space 
                \space\space\space\space\space\space
   (b)}
   \label{} 
\end{subfigure}
\begin{subfigure}[b]{0.62\textwidth}
   \includegraphics[width=1\linewidth, trim= 10mm 70mm 12mm 55mm]{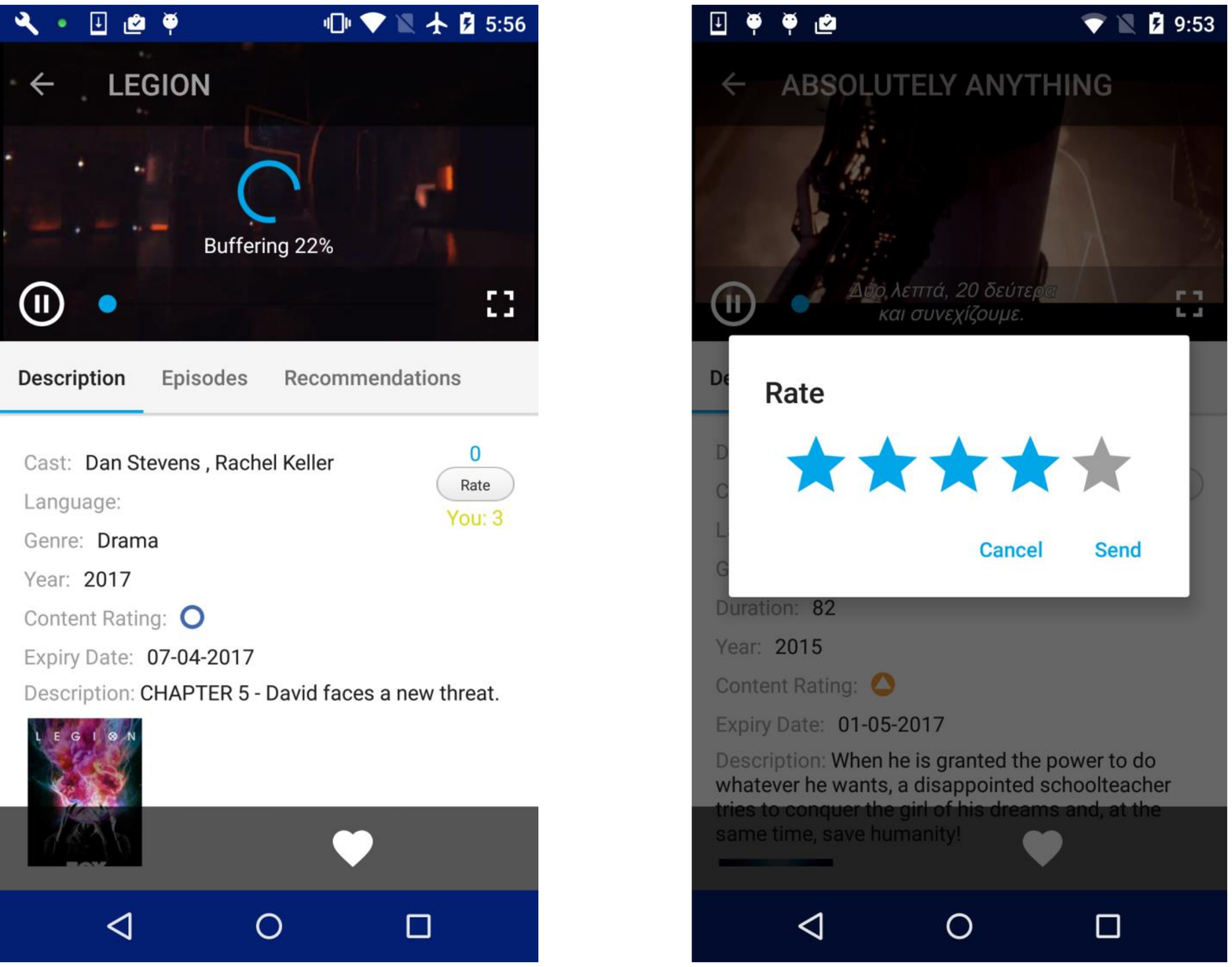}
   \caption{(c) \space\space\space\space\space\space\space\space 
   				\space\space\space\space\space\space\space\space 
                \space\space\space\space\space\space\space\space 
   				\space\space\space\space\space\space\space\space 
                \space\space\space\space\space\space\space\space
                \space\space\space\space\space\space\space\space 
                \space\space\space\space\space\space
   (d)}
   \label{}
\end{subfigure}
\caption{Pilot system (various  screenshots). The video-streaming service that runs on the client with our underluing recommenders. (a) The \textit{Recommended} screen presents the content that is recommended to the user in the form of horizontal lists, ranked by the level of predicted preference. (b) The \textit{OnDemand} screen displays the available content for watching with an order that is customizable by the user (e.g., latest releases or most watched content). (c) The \textit{Play} screen streams the selected content and demonstrates detailed information about it (e.g., movie genres, cast, summary). (d) Users can provide a star rating  for a given content.}
\label{fig:novago}
\end{figure}

When a user selects a video, a \textit{viewing session} is initiated and a new screen appears (referred as "\textit{Play}" screen, shown in Fig.~\ref{fig:novago}~(c) ), which streams the selected video and displays related information (e.g., genres, release year, plot summary, actors). Within this screen, the user can provide a score (i.e., rating) in a five-star scale for the video content at any time (Fig.~\ref{fig:novago}~(d)). A rating is also requested when the viewing session is terminated, either by the user or because of the completion of the video. The duration of the viewing session is defined as the total time spent on the \textit{Play} screen that streams the selected video. A viewing session is characterized as \textit{watched}, if its total duration is 2 minutes or more. Using this 2-minute threshold, we account for situations where the user clicked on a video in order to get detailed information about its content or provide a rating, instead of watching it. 
\par
The client collects information about the user preferences, viewing behavior (e.g., the watching duration), and interaction with the system (e.g., click and scroll events, search queries). It saves this information in an internal SQLite relational database. For handling the communication with the server, the client includes a \textit{back-end interface} module. This module employs a secure HTTP client and the functionality that connects the client to the server using the JavaScript Object Notation (JSON) data-interchange format for communication. During a connection, the client may upload the recorded information to the server.

\par
\begin{figure*}[t!]
\renewcommand{\arraystretch}{1.3}
\hspace{0mm}
\setlength\fboxsep{0pt}
\setlength\fboxrule{0pt}
\fbox{
\centering
\begin{tabular}{ccc}
\includegraphics[width=2.5in, trim= 10mm 0mm 7mm 0mm]{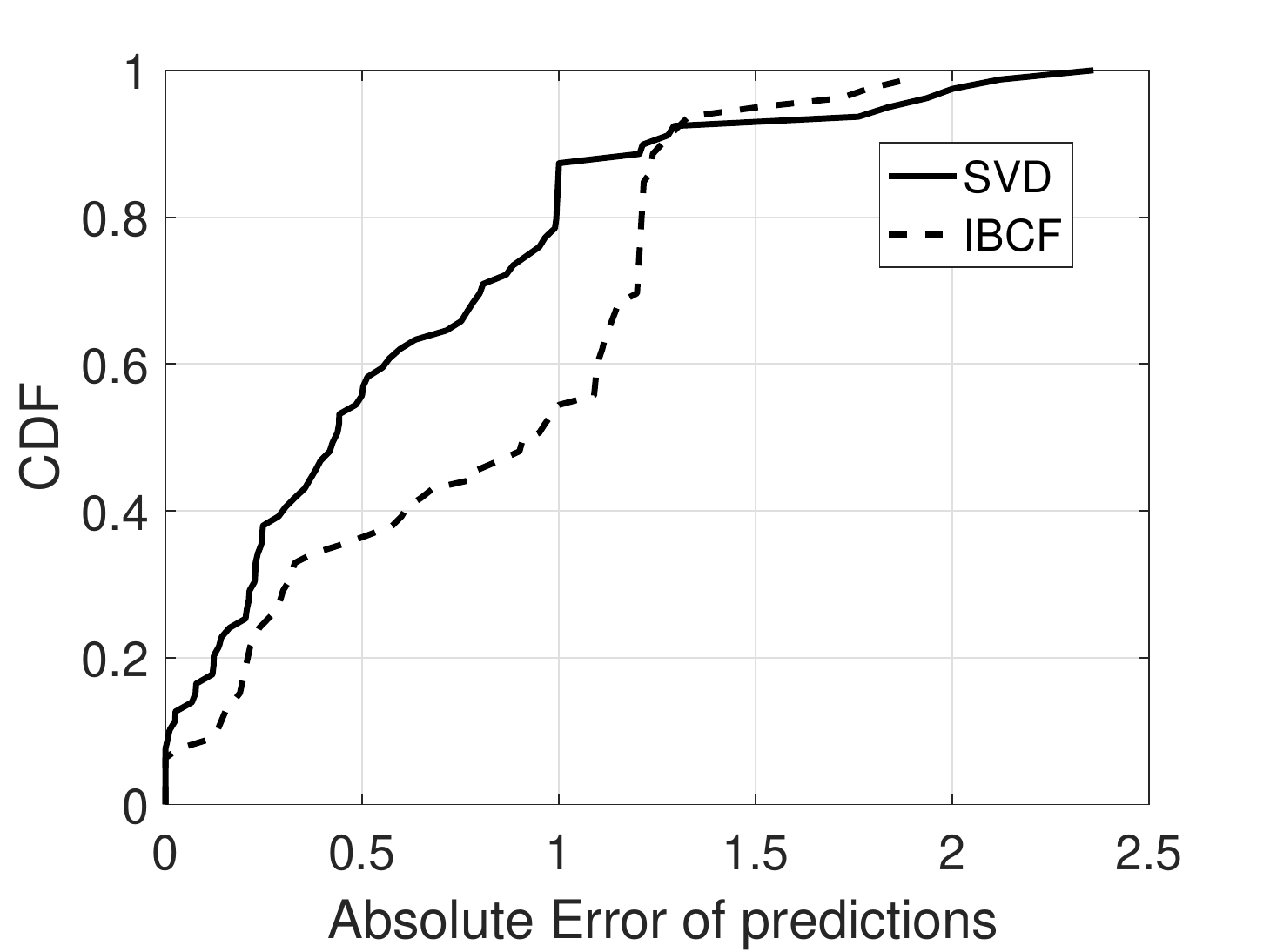}
\put(-110,39){\scriptsize \setlength{\tabcolsep}{4pt}\begin{tabular}{|cccc|} 
\hline
\textbf{Rec} & \textbf{Mean} & \textbf{Median} & \textbf{Std} \\
\hline
\textbf{SVD}	&	 0.59   & 0.43  &  0.54\\
\hline
\textbf{IBCF}	&	 0.79   & 0.95  &  0.51\\
\hline
\end{tabular}}
&
\hspace{7mm}
\includegraphics[width=2.5in, trim= 10mm 0mm 7mm 0mm]{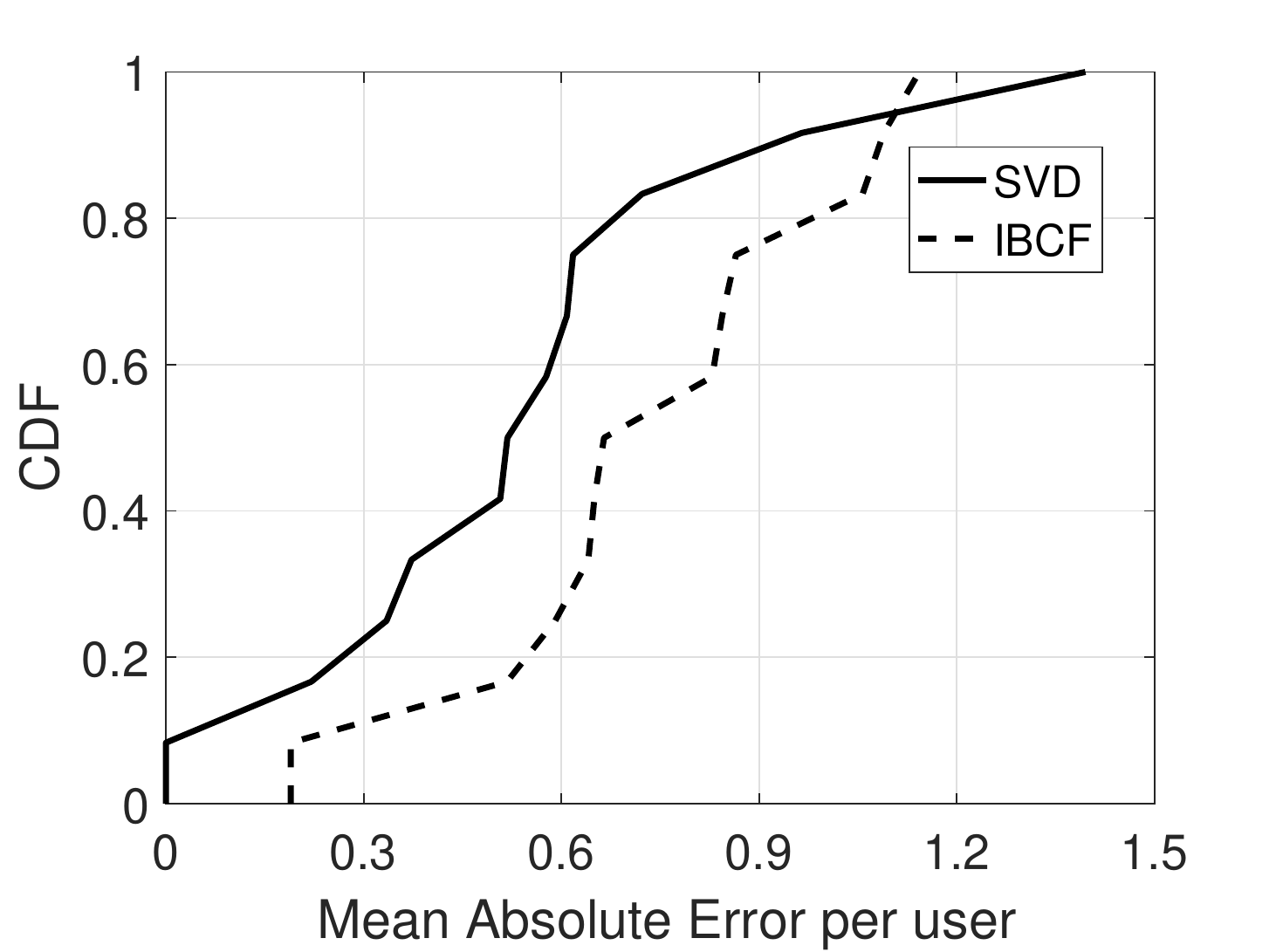}
\put(-110,39){\scriptsize \setlength{\tabcolsep}{4pt}\begin{tabular}{|cccc|} 
\hline
\textbf{Rec} & \textbf{Mean} & \textbf{Median} & \textbf{Std} \\
\hline
\textbf{SVD}	&	 0.57   & 0.54  &  0.35\\
\hline
\textbf{IBCF}	&	 0.75   & 0.74  &  0.27\\
\hline
\end{tabular}}\\
(a) & (b)
\end{tabular}
}
\vspace{-0.7mm}
\caption{Performance of the pilot RS during the field study (a) The absolute error between the score predicted by each algorithm, and the actual rating given from the user. (b) The mean absolute error exhibited for each user by both algorithms.}
\label{fig:cdfs}
\vspace{-3mm}
\end{figure*}

\par

\begin{figure*}[t!]
\renewcommand{\arraystretch}{1.3}
\hspace{0mm}
\setlength\fboxsep{0pt}
\setlength\fboxrule{0pt}
\fbox{
\centering
\begin{tabular}{ccc}
\includegraphics[width=2.5in, trim= 10mm 0mm 7mm 0mm]{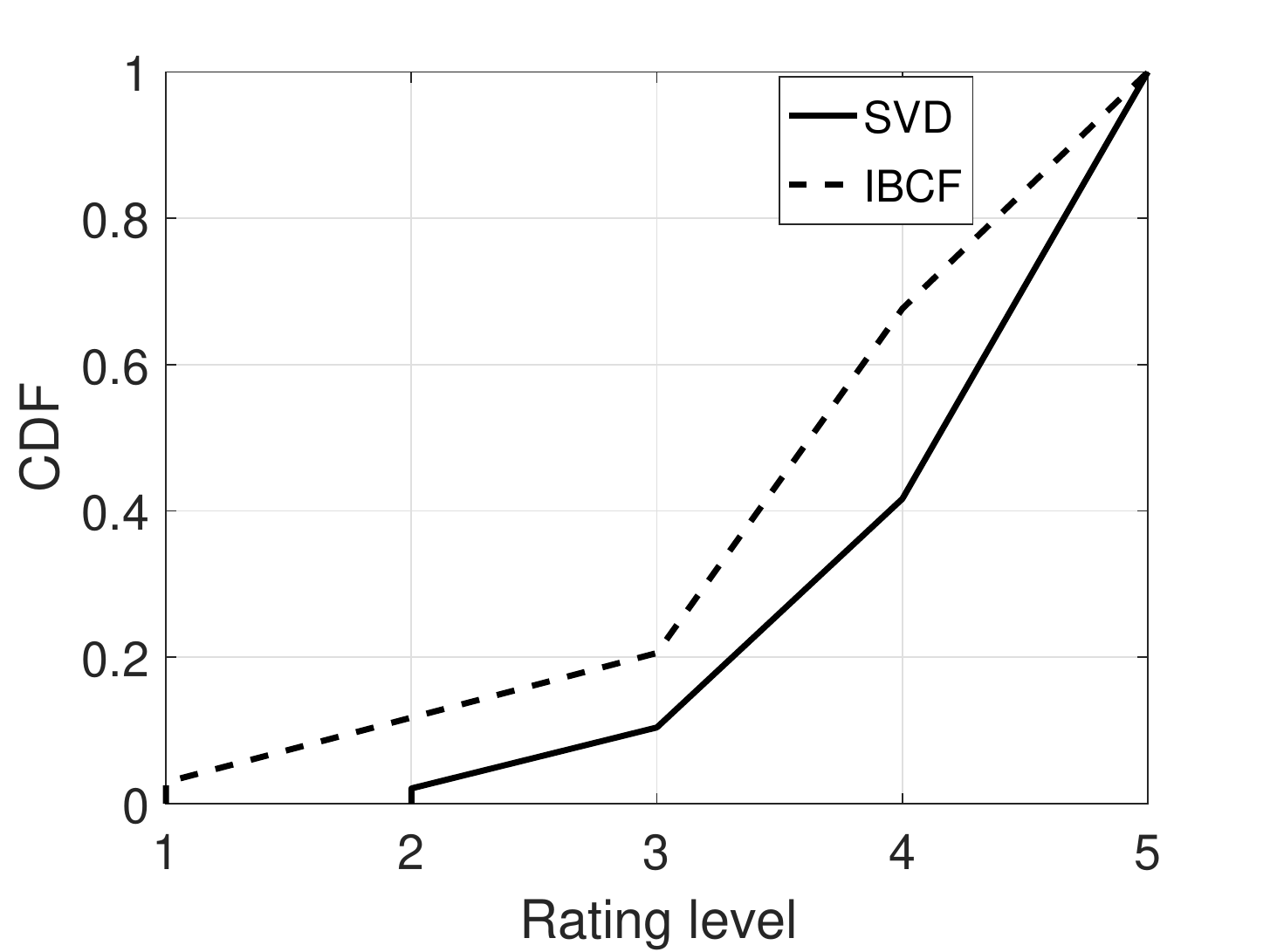}
\put(-167,100){\scriptsize \setlength{\tabcolsep}{4pt}\begin{tabular}{|cccc|} 
\hline
\textbf{Rec} & \textbf{Mean} & \textbf{Median} & \textbf{Std} \\
\hline
\textbf{SVD}	&	 4.45   & 5  &  0.74\\
\hline
\textbf{IBCF}	&	 3.97   & 4  &  1.02\\
\hline
\end{tabular}}
&
\hspace{7mm}
\includegraphics[width=2.5in, trim= 10mm 0mm 7mm 0mm]{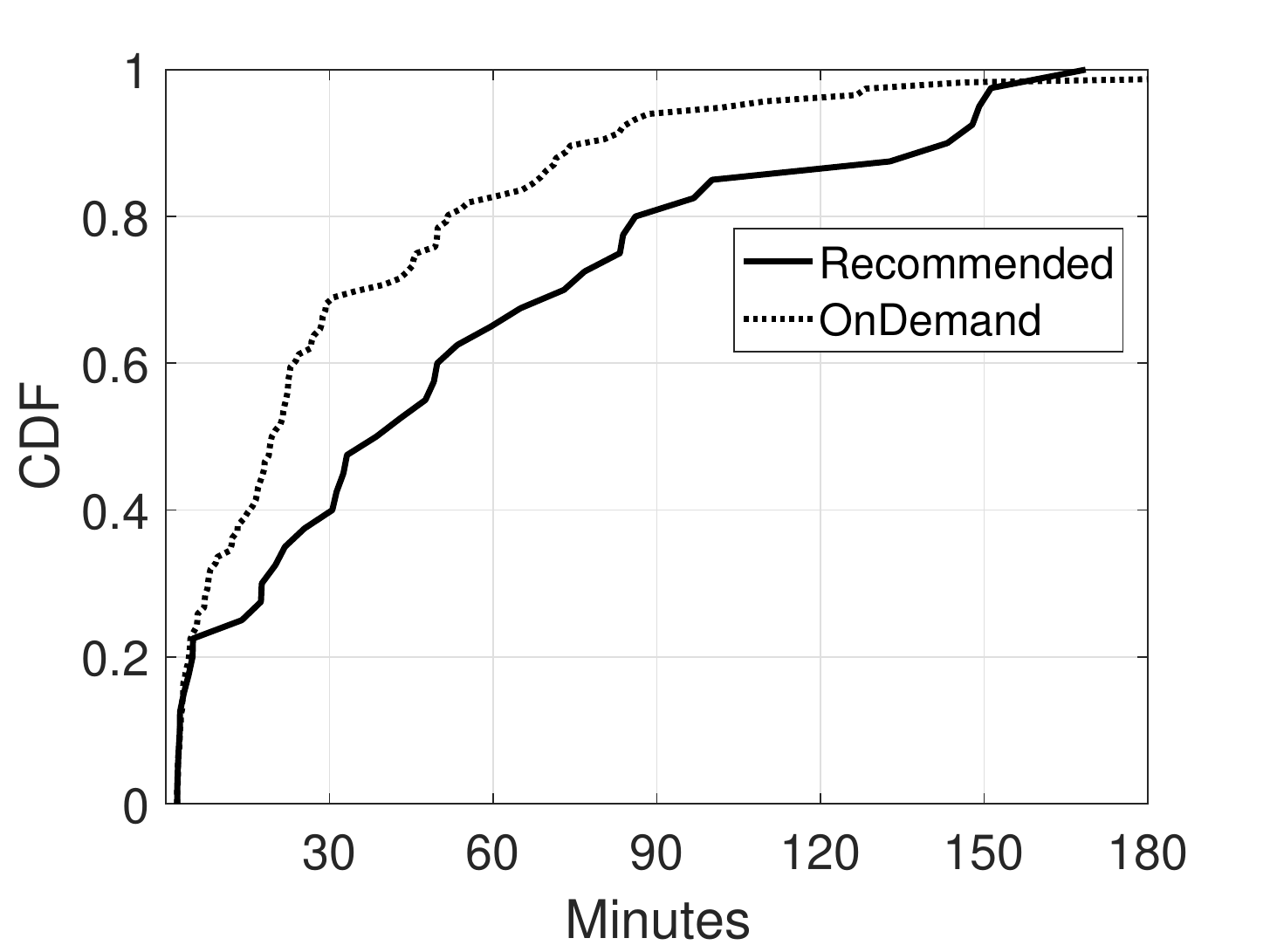}
\put(-140,38){\scriptsize \setlength{\tabcolsep}{4pt}\begin{tabular}{|cccc|} 
\hline
\textbf{Screen} & \textbf{Mean} & \textbf{Median} & \textbf{Std} \\
\hline
\textbf{Recommended}	&	 54   & 41  &  49\\
\hline
\textbf{OnDemand}	&	 33   & 20  &  41\\
\hline
\end{tabular}}\\
(a) & (b)
\end{tabular}
}
\vspace{-0.7mm}
\caption{The performance of the pilot RS during the field study. (a) The ratings given to each algorithm's recommendations. (b) The duration of watched sessions that originated from each screen.}
\label{fig:cdfs2}
\vspace{-3mm}
\end{figure*}

The server, implemented in Java, runs on a Linux virtual machine. The JSON-formatted data received from the clients are stored in a PostgreSQL object-relational database. The recommenders are trained periodically, once a day, by  taking into account the new ratings and predicting the ratings that have not been provided by each user. The predicted ratings of each algorithm are ranked in descending order, forming the \textit{SVD} and \textit{IBCF} \textit{recommendation lists}, so that the content which is predicted as the most preferred appears at the top positions of the corresponding lists. The personalized recommendation lists are built \textit{offline} and stored in the database, in order to be available upon a user login to the service in a timely manner. 
\par
The user interface of the video-streaming client was extended for the purpose of appropriately presenting the recommendation lists  on the "Recommended" screen, which is shown in Fig.~\ref{fig:novago} (a). The "Recommendations A" correspond to recommendations produced by the SVD, while the "Recommendations B" are derived from the IBCF. In addition to the aforementioned personalized recommendation lists, we included a {\em non-personalized} list ("Recommendations C") which displays recently-added content to the catalog. The users can also discover content through the "OnDemand" screen (Fig.~\ref{fig:novago} (b)), which presents all available content and allows users to customize the display, in terms of the type of content to be shown (e.g., only TV series) and how it is ranked (e.g., based on the release year).  The GUI only displayed the recommendation lists A, B, and C, without providing any technical information about the underlying algorithms to the user. The order of these three recommendation lists, in terms of how they were displayed on the user device, remained fixed throughout the field study.
\par

\subsection{Field study}

To evaluate the performance of the pilot RS, we performed a field study with volunteer subscribers in the operator's production environment. The subjects were asked to provide preference information about the content, in order to receive personalized recommendations. We assessed the prediction accuracy of both algorithms and examined whether the provided recommendations improve the user engagement. The subjects were not provided any information about the recommenders to avoid any bias.
\par
During the field study that lasted from 28 February until 17 April 2017 (7 weeks), 30 volunteer customers of the service participated by viewing content and providing ratings. The collected dataset includes 387 ratings given by 27 users to 182 distinct movies, and 187 watched sessions originated from 22 users. To estimate the accuracy of the algorithms, we focus on the ratings that the users have given to videos for which a predicted score had already been calculated (79 out of the total 387 ratings). Fig.~\ref{fig:cdfs} (a) shows that the SVD outperforms the IBCF in terms of mean and median absolute error with values 0.59 and 0.43, respectively. It is noticeable that the IBCF has a median error of 0.95, which is approximately twice as large as the one of the SVD. To further investigate the performance of the SVD across users, we calculate the mean absolute error (MAE) both algorithms exhibit for each user (Fig.~\ref{fig:cdfs} (b)). For the large majority of users, the SVD achieves more accurate predictions than the IBCF. The median MAE obtained by the SVD is 0.54, compared to 0.74 achieved by the IBCF. The outcome of this analysis is consistent with results reported in \cite{ekstrand2015letting,ekstrand2011rethinking,guo2015librec}. Kluver \textit{et al.} \cite{kluver2014evaluating} also indicate that in the case of a small number of ratings, the IBCF may perform poorly. 
\par
To determine the user satisfaction towards the recommended content, we examined the ratings that are given to the recommendations:  the SVD recommendations tend to receive higher ratings, than the ones derived from the IBCF, which suggests that the first performs more efficiently in terms of recommending highly preferred content, than the latter (Fig.~\ref{fig:cdfs2} (a)). More specifically, the median (mean) rating for the SVD recommendations is 5 (4.45), respectively, while for IB 4 (3.97), respectively.
\par
In addition to the prediction accuracy, the watching duration is another metric for assessing the user satisfaction with the recommendations. Since the recommendation task aims at aiding users in discovering content to watch, an effective RS recommends content users opt for watching. To examine whether or not the pilot RS leads to higher consumption of content, we analyzed the duration of the watched sessions that originated from the "Recommended" and the "OnDemand" screens. The median duration of the watched sessions initiated from the recommended content has a value of 41 minutes, which is twice as large as the median duration of the ones started from the "OnDemand" screen (20 minutes) (Fig.~\ref{fig:cdfs} (b)). That is, the viewing session of recommended content is longer compared to the one selected from other sources.

\begin{figure}[t!]
\renewcommand{\arraystretch}{1.4}
\centering
\includegraphics[width=5in]{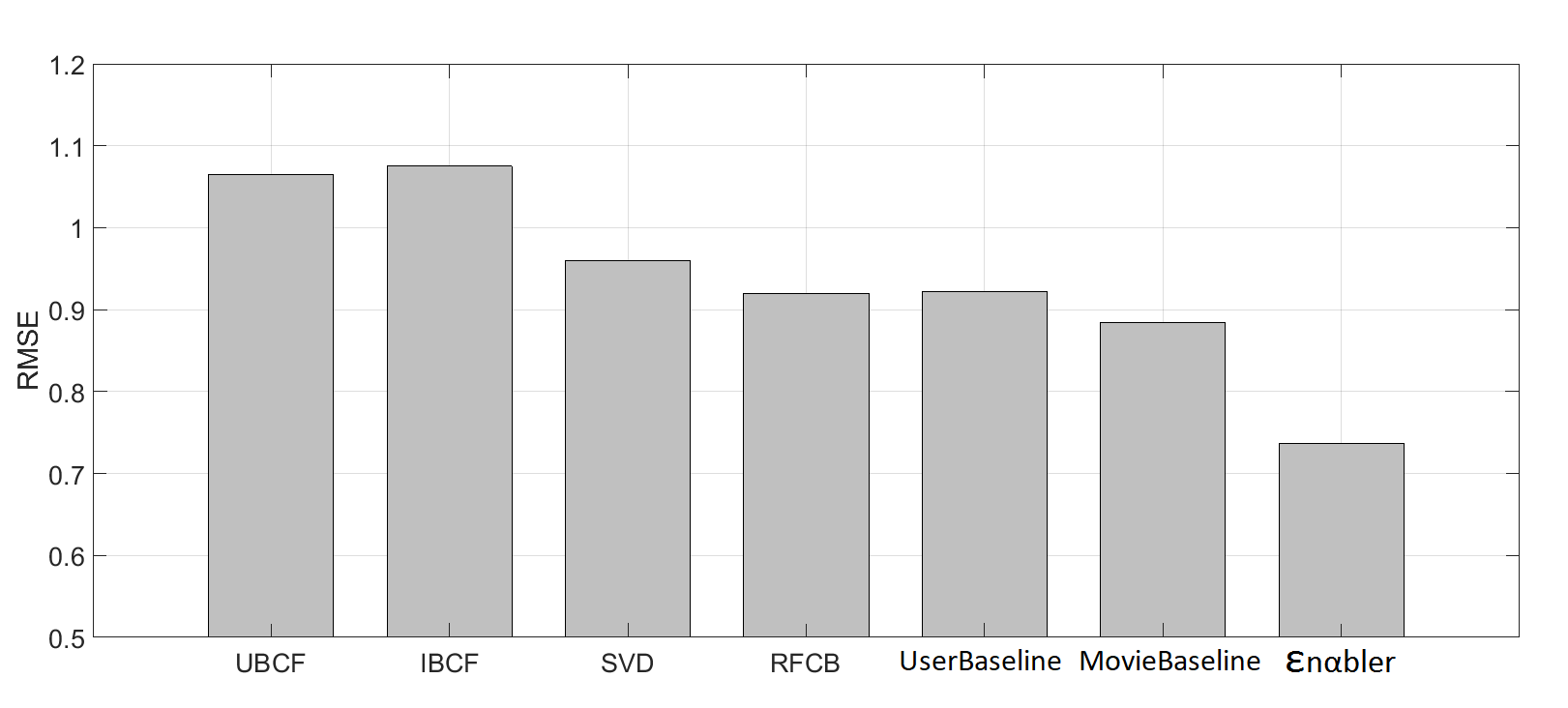}
\caption{Post-phase, offline evaluation of the \enabler, compared to several recommendation algorithms using data collected during the field study.}
\label{fig:forthnet_indiv}
\end{figure}

\subsection{Offline evaluation of \enabler}
Based on the ratings collected during the field study, we performed a post-phase, offline evaluation of the \enabler. We aimed to examine whether an efficient combination can be achieved, in order to further improve the prediction accuracy in comparison to the one achieved by each recommender individually. Due to the small number of samples in the field study dataset, we performed a 20-fold CV for assessing the accuracy of each individual algorithm, meaning that each one of them is trained using 19 folds of subsamples and tested on the remaining fold. This procedure is repeated 20 times, so that each subsample has been used for testing. For evaluating the performance of the \enabler, the 20-fold nested CV is employed, which uses 20-fold CV for the tester layer, and 19-fold CV for the blender layer.     
\par
The recommenders that are employed and tested are shown in Fig.~\ref{fig:forthnet_indiv} along with their average RMSE. The collaborative filtering algorithms (e.g., SVD, UBCF and IBCF) exhibit a relatively poor performance under the collected dataset. Since only 5 users have provided at least 20 ratings, there is no adequate information about the user preferences. For this reason, the collaborative filtering algorithms fail to capture the user preferences sufficiently. Among the latter, the SVD exhibits the most accurate predictions (0.9599 RMSE), complying with the results in \cite{kluver2014evaluating} which shows that the SVD is more accurate for users with a few ratings than the KNN techniques. The RFCB, which builds a profile for each user based on the genres of the movies they have rated, outperforms all the collaborative filtering algorithms in terms of accuracy (0.9198 RMSE), since it deals more efficiently with the cold-start problem. The UserAverage and the MovieAverage are meta-features estimated for each user and movie in the dataset, reporting the average rating a user provides and a movie receives, respectively. These meta-features can be directly used to predict a rating. Kluver \textit{et al.} \cite{kluver2014evaluating} suggest that in cases of users with limited information about their preferences, these baselines can provide more accurate predictions than the CF algorithms, also consistent with our analysis. In particular, the MovieAverage algorithm has the lowest RMSE (0.8844) among the individual recommendation algorithms.
\par
The average RMSE values of each individual recommendation algorithm according to a 20-fold CV was estimated. The \enabler~outperforms all individual algorithms (Fig.~\ref{fig:forthnet_indiv}). More specifically, the \enabler~achieves a RMSE of 0.7372 which is a substantial improvement (16.64\%) against the MovieAverage, which is the best performing individual recommendation algorithm. Additionally, the RMSE decreases by 19.85\%  and 23.20\% compared to the RFCB and the SVD algorithms, respectively. These results reflect the benefit of blending. The \enabler~efficiently combines various recommendation algorithms, which perform relatively poorly when used individually, to provide more accurate final predictions.  
\par

\begin{figure}[h]
\renewcommand{\arraystretch}{1.3}
\centering
\includegraphics[width=4.4in]{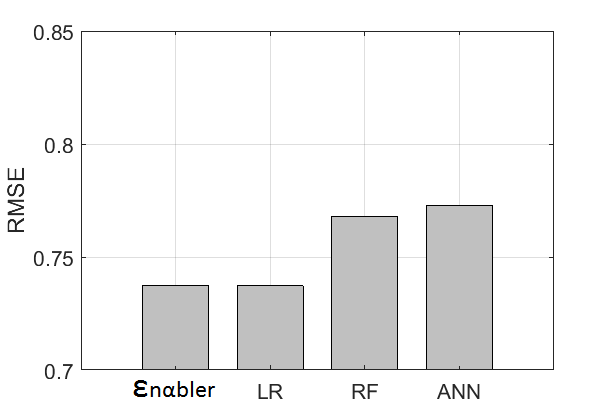}
\put(-270,119){\scriptsize \setlength{\tabcolsep}{10pt}
\footnotesize
\begin{tabular}{|c|c|} 
\hline
\textbf{Algorithm} & \textbf{RMSE}\\
\hline
\enabler & \textbf{0.7372}  \\
\hline
LR  & \textbf{0.7372}  \\
\hline
RF & 0.7679 \\
\hline
ANN & 0.7730 \\
\hline
\end{tabular}}\\
\caption{Post-phase, offline evaluation of the performance of each blending algorithm.}
\label{fig:enabler_nested_forth}
\end{figure}

We further investigated the performance of each blending algorithm separately with the Nested CV. As shown in Fig.~\ref{fig:enabler_nested_forth}, the linear regression exhibits the lowest RMSE (0.7372) and is selected as the best blending model from the \enabler, followed by the random forest and the ANN (0.7679, and 0.7730, respectively). The performance under the blending algorithms is better than when individual recommendation algorithms are used separately. Although the ANN achieves the most efficient combination in the Movielens dataset, it fails at discovering valuable relationships among the individual algorithms in the field study dataset, which can be better identified by the linear regression. These results further illustrate the advantage of the \enabler, which can automatically determine the most efficient blending technique, based on the input dataset. 
\clearpage
\section{Conclusions and future work}
In this work, we developed the \enabler, a hybrid recommendation algorithm that employs a number of ML algorithms for combining a diverse set of widely-employed and state-of-the-art recommenders, and automatically selects the best according to the input. It outperforms state-of-the-art recommendation algorithms, such as the AutoRec \cite{sedhain2015autorec} and SVD \cite{koren2009matrix}, achieving a RMSE of 0.8206, compared to 0.827 and 0.845, respectively, on the benchmark Movielens 1M dataset. 
\par
The web-based pilot RS based on the \enabler, developed for a mobile video-streaming service provided by a major Greek telecom operator, exhibits a very good performance. 
The pilot RS appears to aid users in discovering content to watch, given that the median watching duration of the recommended content is double that of the non-recommended (41 minutes, and 20 minutes, respectively). The offline post-analysis of the \enabler, using the collected ratings of the pilot study, reported a significant improvement in the prediction accuracy, compared to popular recommenders (RMSE improvement higher than 16\%). 
\par
Recurrent neural networks can be applied to identify the contextual and temporal features with dominant predictive power on user engagement. By analyzing the user preference time-series, we can
detect changes in trends, e.g., towards the preferred content (\textit{change-point} problem \cite{pettitt1979non}) and determine when a system
retraining is required. This can potentially improve the computational
complexity and scalability of our system, and the overall quality of recommendations. Moreover, the active querying of a small number of selected users (i.e., targeted queries) can further improve he overall accuracy \cite{ruchansky2015matrix}.
\par
It is part of our on-going research the incorporation of information about the user viewing behavior (e.g., watching duration, pause/skip events during playback) and user interaction with the system (e.g., time spent reviewing recommendations), for inferring the user preference/engagement with respect to certain movies, especially when a rating is not available.   
\par
\par
Our long-term objective is the generalization of our system to be also applied in other domains, with an emphasis on the personalized health care and medicine. The provision of personalized treatment plans, taking into account individual information associated with each patient, can potentially improve the quality of care, address real-time constraints, enhance the educational process of medical students, and decrease medical costs.

\section*{Acknowledgement}
This work was partially supported by Forthnet S.A. The authors would like to thank the team of Forthnet S.A., and in particular, Ioannis Markopoulos and George Vardoulias, for their help in deploying the system and performing the field study.

\bibliographystyle{plainnat}
\bibliography{references}



\end{document}